\documentclass[aps,twocolumn]{revtex4-1} 
\usepackage{amsmath}
\usepackage{epsfig}
\usepackage{graphicx}
\usepackage{color}
\usepackage{amssymb}
\usepackage{enumerate}
\begin{document}
\title{Interwoven limit cycles in the spectra of mass imbalanced many-boson system}
\author{W. De Paula$^1$}
\author{A. Delfino$^2$} 
\author{T. Frederico$^1$}
\author{Lauro Tomio$^{1,3}$} 
\affiliation{
$^1$Instituto Tecnol\'ogico de Aeron\'autica, 12228-900, S\~ao Jos\'e dos Campos, SP, Brazil\\
$^2$Instituto de F\'\i sica, Universidade Federal Fluminense, 24210-310 Niter\'oi, RJ, Brasil.\\
$^3$Instituto de F\'{\i}sica Te\'orica, Universidade Estadual Paulista, 01140-070, S\~ao Paulo, SP, Brazil
}
\date{\today}
\begin{abstract} 
The independence between few-body scales beyond the van der Waals universality is demonstrated for
the extreme mass-imbalanced case of a specific many-boson system. This finding generalizes   
the scaling properties of universal tetramers to a broader class of heterogeneous few-boson systems.
We assume two heavy atoms interacting with $(N-2)-$lighter ones at the unitary limit, using a particular case where 
no interactions are active between identical particles, by investigating 
the interwoven spectra of this many-body system for an arbitrary number of light bosons. 
A large mass-ratio between the particles allows us to treat this $N-$body system analytically, by solving 
an effective inverse-squared long-range interaction which is stablished for the two heavy bosons. 
For a cluster with $N-2$ light bosons ($N\ge 4$), we discuss the
implications of the corresponding long-range potentials associated with different subsystem thresholds, implying in  
independent interwoven limit cycles for the correlation between the energies of excited $N-$body system.
 Our study with extreme mass-imbalanced few-boson bound states
provides a fundamental understanding of the scaling behavior of their interwoven spectra.
The novel insights enlarge the well-known Efimov physics paradigm and show 
the existence of different limit cycles, which could be probed by new experiments.
\end{abstract}
\maketitle

\section{Introduction}\label{introduction}
The existence of a universal correlation between the binding energies of successive four-boson bound states (tetramers), 
for large two-body scattering lengths, was verified by solving the corresponding four-body Faddeev-Yakubovsky (FY) 
formalism in Ref.~\cite{2011hadizadeh}. The results obtained are related to the existence of additional 
scales~\cite{AdhPRL1995,YamaEPL2006} not constrained by the well-known three-body Efimov physics~\cite{1970efimov}. 
The correlations verified in Ref.~\cite{2011hadizadeh} are further explored in Ref.~\cite{2013hadizadeh} in an 
application to an ultracold gas of cesium atoms close to broad Feshbach resonances,
by considering the shifts in the four-body recombination peaks, due to an effective range correction to the zero-range 
model close to the unitary limit.
However, by considering a bound system with more than three particles in the unitary limit, a challenge was in establishing 
a more simplified description (possibly relying in some analytical procedure) to confirm the emergence of an independent 
new scale when a new particle is being added to the system. In a more general context of a many-body system, the results 
for the independent spectra could be verified by the interwoven between the corresponding limit cycles. 

In the context of cold atoms such independence between few-body scales is beyond the van der Waals 
universality~\cite{BerPRL2011,ChinArXiv2011,WangPRL2012,SchmidtEPJB2012,NaidonPRL2014,NaidonPRA2014,WangNPHYS2014}.  
The van der Waals universality is specific to the case of  atoms interacting by the corresponding inverse power law
potential \cite{WangNPHYS2014}, and it is verified  close to a broad Feshbach resonance dominated by the open channel.  
It is due to the  separation between the ranges of about one nanometer of the true short-range few-atom
chemical  potentials and the van der Waals length of several nanometers. At such distances, the atoms
have an effective repulsive force which prevent them to explore shorter length scales in order to be sensitive to the details of the 
interatomic potentials. Therefore, the van der Waals length is the single short range scale that controls and determines the few-atom physics. 
This is supported by several experiments that have investigated the position of three and four-atom recombination peaks 
(see e.g. \cite{IncaoJPB2018}).
While this would suggest a general universality for all systems with arbitrary number of bosons 
(see e.g.~\cite{KievskyPRA2014,2011StecherPRL,2017KievskyPRA}), a recent work \cite{YanPRA2015} has shown that   
the ground-state energy and structural properties for larger clusters of identical bosons interacting via a two-body zero-range force regulated 
at finite range are not universally determined by the three-body parameter, as it was found theoretically that the results will depend on the specific 
form of the three-body regulator. As already found in the case of Brunnian (Borromean with arbitrary number of particles) systems with identical 
bosons \cite{YamaPRA2010},  the binding energies at unitarity show a large variability when obtained with short-range interactions.
In a recent  review~\cite{GreeneRMP2017}, the reader can found an updated discussion on the issues raised by the existence of  scales beyond 
the three-body one in few-boson systems.

Recent experimental investigations on the predicted universality of Efimov states across broad and narrow Feshbach 
resonances~\cite{2014chin,2017chin} with the Lithium-Caesium  ($^6$Li-$^{133}$Cs) mixture are opening a new window of 
opportunities to test the independence between the few-body scales, where the so called van der Waals universality may be broken.
A single channel prescription is not enough to describe the position of the three-body recombination peaks, as the results so far
obtained are evidencing a dependence of the position of the  Efimov resonance  on the Feshbach resonance strength, with  
a clear departure from the universal prediction for the narrow Feshbach resonance.
Near a narrow Feshbach resonance, where a single channel description is poor, it is natural to presuppose that, 
beyond the expected large variations in the two-body scale (namely, the atom-atom scattering length, $a$), the three-body parameter can also move 
as suggested by the observations of resonant recombinations in the $^6$Li-$^{133}$Cs$_2$ experiments~\cite{2017chin}. 
On theory grounds, such possibility can occur as three- and even four-body potentials are induced when the atomic trap set-up 
is tuned to be close to a narrow Feshbach resonance, such that 
the single channel description has to be supplemented by these induced potentials in the open channel \cite{YamaEPL2006}.
When this happens, one can disentangle the effects of the independent three, four and more body scales, which indeed are called to 
justify a thoroughly discussion of the independence of these scales for trapped atoms  in \cite{YamaEPL2006}. 

In brief,  the origin of the induced few-particle interactions can be understood by starting with the Feshbach decomposition of the 
Hilbert space in open channels ($P$-space)  and closed channels ($Q$-space) with $P+Q=1$. 
In the atom-atom Feshbach resonance, the $Q-$space represents the pair states in the potential well where it is bound,  
while the $P$-space contains the pair states in the lower potential well, where  the open channel wave function meets 
asymptotically the continuum. To be concrete, the projectors $P$ and $Q$ will represent different spin states of the atom-atom pair, 
which are close to a Feshbach resonance (see e.g. \cite{WangNPHYS2014}).
For few-body systems, for example the ones constituted by three and four particles, the three- and four-body  effective potentials appear in the 
single channel description, respectively, when  the pair of particles virtually propagates in the $Q-$space, interacting with the 
spectators. Such discussion is detailed and illustrated in \cite{YamaEPL2006}.

The above mentioned example considers one Feshbach resonant pair interaction in the three atom system. 
 By reducing the three-body coupled channel problem to the open single channel one, 
 the effective Hamiltonian acting on this channel contains 
 an effective potential, which has a non-connected three-body part corresponding to the resonant pair interaction 
and a  connected three-body part that corresponds to an effective three-body interaction with intensity depending on the 
properties of the Feshbach resonance. 
In particular, the strength of the effective potential is enhanced for a narrow resonance, as the coupling between open 
and closed channels is larger  in this case~\cite{WangNPHYS2014}.
Note that, in the region where the effective three-atom potential is attractive,
the length scale is larger than the van der Waals length. 

The attraction dislocates the effective repulsive barrier to distances larger than the van der Waals radius $(\ell_{vdW})$,
with the characteristic length scale increasing with respect to $\ell_{vdW}$. This implies in  
a dislocation of the position of the resonance in the three-body recombination towards larger  
absolute values of the scattering length. Indeed, this behavior was verified in recents experiments, as reported in 
Ref.~\cite{2017chin}.
By extending such example to systems with more particles, one should expect that, 
through Feshbach resonance mechanisms, not only the two-body scattering length is disposable to be tuned,
but also the short-range scales related to three, four and more particles. 
In the presence of spectator particles, if the excitation of the Feshbach resonance is turned off, this example will reduce 
to the one discussed in Ref.~\cite{jonsell}.
 
  In short, near a narrow Feshbach resonance, the induced few-body forces in the open channel can drive independently 
the corresponding physical scales. Then, by restricting our example to a four-particle system, observables such as 
the position of the resonance in the recombination rates, or the scattering lengths for atom-dimer, atom-trimer
and dimer-dimer, are not constrained only by the van der Waals length, as other larger few-atom length scales 
can play a role. 

In view of the above considerations, we are motivated to search for a generalization of previous findings obtained  
in the case of a four-boson system with zero-range two-body interactions~\cite{2011hadizadeh}, in which 
a four-body scale was found necessary when the two- and three-boson scales are fixed, leading to the prediction of
a new limit cycle for the four-boson system. 
   Such limit cycle has a different geometrical ratio between the four-body energies in the unitary limit, as compared 
   to the usual Efimov ratio, when the trimer energy is smaller compared to the tetramer energy. This suggested that 
   the long-range effective potential for the tetramers has a strength different from the trimer one. In addition, the 
   four-body state can be moved by a four-body short range interaction, while keeping the universal correlation between 
   two successive tetramer states. 
   Narrow Feshbach resonance may allow to disentangle the trimer and tetramer binding energies. 
   For more particles, it was suggested that a new experimental information is required for each new boson added to the 
   system~\cite{AdhPRL1995}. Narrow Feshbach resonances, 
   which may include effects from many-body forces when the dynamics is reduced to a single channel, 
   can  be effectively described by  few-body scales that move independently. 
   In the particular case, when the interaction is dominated by the open channel, with the single channel description well established 
   and only two-body forces being manifested, all short range scales should be determined by the van der Waals length. 
   Considering  such quite exciting possibilities for narrow Feshbach resonances, it is timely and demanding a study on 
   the interwoven and independent cycles, which can emerge in the spectrum of a many particle system.
In addition, the interest in few-body physics and the corresponding scaling behavior of the observables, following studies
presented in Refs.~\cite{2006Kraemer,2017Incao}, is further strengthened by recent experimental observations of three-photon bound states 
in a quantum nonlinear medium, where three-photon bound states are viewed as photonic solitons in the quantum regime~\cite{2018Liang}. 
It was also pointed out in this reference that strong 
effective $N-$body forces in larger photonic molecules and clusters can allow studies 
which are not possible to be realized with conventional systems.
 
Furthermore, as considered in Ref.~\cite{petrov1} for three-bosons 
in the vicinity of a narrow Feshbach resonance, the so-called ``energy-dependent scattering length'' $a(E)$
(as derived from the effective-range expansion) has  
an effective-range correction $R^*$ inversely proportional to the width of the resonance. 
So, for a narrow Feshbach resonance, this range $R^*$ can be a relevant parameter to be taken into account.
It is also noticeable that, within the Born-Oppenheimer (BO) approximation for a heavy-heavy-light system,
the three-body potential acquires a Coulomb character departing from the Efimov inverse-square behavior
at distances $R\ll R^*$~\cite{petrov2}. This 
Coulomb-like character at short distances was also verified in  Ref.~\cite{2018Shalchi} for the scattering of the 
heavy particle by the dimer formed the heavy-light system, near the unitarity.

However, besides being quite relevant to consider a more realistic model which encapsulates the range $R^*$ 
in the physics of deeply bound and more compact systems, 
in the present work we are mainly concerned with the tail of the effective potential 
which is not affected by the effective range in a significant way for the situation that $ R^* \ll R \ll |a|$. 
Our aim is to study weakly-bound $N-$body systems constituted by two-heavy bosons and $N-2$ light ones, close to the $N-1$ threshold ($N\ge 4$) and at the unitarity,
and the states that are characterized by sizes much larger than $R^*$.  
It is worthwhile to to point out \cite{petrov2} that in the case of the heavy-heavy-light systems at distances $R\sim R^*$ the heavy-heavy wave function 
can be matched with the Efimov-like wave function, namely the one living in the long range potential inverse square potential, with the
 three-body parameter determined by $R^*$. In this situation three-body observables depend only 
 on $a$  and $R^*$ \cite{petrov2}.
For the deeply bound systems a departure from the Efimov-type  scaling is expected when their  sizes $\sim R^*$, 
this interesting case is beyond the present investigation.
By using the BO approximation, where the light-heavy system is providing the interaction for the 
heavy-heavy system, we look for a simplified approach considering only the tail of the BO effective potential from 
which the main physics aspects will emerge associated with the few-body scales in correspondence with
 the number of bosons $N\ge 4$ near the unitarity. The present analysis will also support previous studies 
 on the four-body scale in more involved numerical approaches using the FY formalism~\cite{2011hadizadeh,YamaEPL2006}.
 
For our task in evidencing the existence of independent scales in a few-boson system, 
the well-known adiabatic BO approach, applied to a low-energy system with two-heavy and 
one light particles~\cite{1979fonseca,2011bhaduri}, is extended to an $N-$body system with an arbitrary number 
of identical $(N-2)-$light particles. So, for $N\ge 4$, in order to be strictly valid the adiabatic approach, we consider 
the two identical heavy particles ($\alpha=$ 1 and 2, with masses $m_\alpha$) interacting with the light 
particles ($\beta=$ 3 to $N$, with masses $m_\beta \ll m_\alpha$) near the unitary limit. 
Within this procedure, an effective two-body interaction emerges for the two-heavy particles.
The limits of validity of the adiabatic approach is being verified in case of a three-body system, by  comparing with
exact numerical approaches for different two-body interactions and mass ratios. 
Next, the implications are discussed in terms of new interwoven limit cycles.

\section{Born-Oppenheimer approximation} 

\subsection{Two-heavy and $(N-2)$-light boson system}
Here we briefly describe a generalization of the approach presented in Refs.~\cite{1979fonseca,2011bhaduri}
for the case of a many-body mixture with two-species of particles, two-heavy and $(N-2)$-light ones,
we define the corresponding coordinates as ${\bf x}_1, {\bf x}_2$ for the two heavy particles, being ${\bf x}_j$
$(j=3,4,..., N)$ for the $(N-2)$-light particles. 
Next, we consider the minimal condition for the interactions, such that the identical particles are not interacting
between each other, remaining only the heavy-light interactions.
Within this condition,
we define the relative coordinates as 
${\bf R}=({\bf x}_1-{\bf x}_2)$ and  ${\bf r}_{j=1,2,...,N-2}=\left({\bf x}_{j+2}-\frac{{\bf x}_1+{\bf x}_2}2\right)$. 
The corresponding Schr\"odinger equation is given by
\begin{eqnarray}
H \Psi =\left[ -\frac{\hbar^2}{m_\alpha}{\nabla^2_R}+V_0(R)+ \sum_{j=1}^{N-2}H_j
\right]
\Psi,\label{eq1}
\end{eqnarray}
where $\Psi\equiv \Psi ({\bf r}_1,{\bf r}_2,...,{\bf r}_{N-2},{\bf R})$ is the total wave function, 
$V_0$ is the potential between the two-heavy particles, and 
$H_{j}$ is a three-body Hamiltonian corresponding to the interaction between the two heavy particles
with each light particle ${j}$. $H_{j}$ is given by
\begin{equation}
H_j = -\frac{\hbar^2}{2\mu_{(2\alpha)\beta}}{\nabla^2_{\bf r_j}} + 
\sum_{i=1}^{2}V_{i}\left(\left|{\bf r}_j+(-1)^i\frac{\bf R}{2}\right| \right),\end{equation}
 where $\mu_{(2\alpha)\beta}\equiv 2m_\alpha m_\beta/(2m_\alpha+m_\beta)$ is the reduced mass 
 for the $\alpha\alpha\beta$ system and $V_i$ is the interaction for the heavy-light system.
 
 The heavy particles should move much slower than the light one, in such a way that we can apply the 
 Born-Oppenheimer approximation. Within this limit, the total wave function can be decomposed as 
 $$\Psi\equiv \Psi ({\bf r}_1,{\bf r}_2,..., {\bf r}_N,{\bf R})= \phi({\bf R})
\prod_{j=1}^{N-2}\psi_{\bf R}({\bf r}_j), $$ 
  where ${\bf R}$ is a parameter in  $\psi_{\bf R}({\bf r}_j)$. 
   { 
   Within our assumption that all the $N-2$ light particles interact in the same way with the heavy particles,
 ${\cal E}_{N-2}(R)\equiv(N-2){\cal E}(R)$ will be  
 the effective potential for the two heavy particles:   
{\small \begin{eqnarray}
&&\left[
-\frac{\hbar^2}{2\mu_{(2\alpha)\beta}} \nabla^2_{r_j} + 
\sum_{i=1}^{2} V_i \left(\left| {\bf r}_j+(-1)^i \frac{\bf R}{2} \right|\right)
- {\cal E}(R)\right]
\psi_{\bf R}({\bf r}_j)=0\nonumber\\
&&\left[-\frac{\hbar^2}{m_\alpha}{\nabla_R^2} + V_0(R)+{\cal E}_{N-2}(R)\right]\phi({\bf R})=  
E_{N}\phi({\bf R}),
\label{eq2}\end{eqnarray}
}    
with $E_3$ being} the energy solution for the system with two-heavy and one light boson.
As the asymptotic behavior of  ${\cal E}(R)$ is not affected by $V_0({\bf R})$, we can assume 
$V_0({\bf R})=0$ within our purpose. For the light-heavy particles one can take short-range separable 
interactions, with $V_1$ and $V_2$ having the operator form $\lambda |g\rangle\langle g|$.
In this way, the light-heavy particle system can easily be solved in momentum space by considering 
{  simple separable interactions with Yamaguchi form-factors such as 
$g(p)\equiv 1/(p^2+\beta^2)$. For another choice of form-factor, with two parameters allowing to
reproduce low-energy phase shifts together with the corresponding dimer energy, see 
Ref.~\cite{2007-shepard}.
} 
Further, it is assumed a shallow 
bound state, $-\hbar^2/(2\mu_{\alpha\beta} a^2)$, where $\mu_{\alpha\beta}$ is the reduced
mass and $a\equiv a_{\alpha\beta}$ the { 
light-heavy scattering length. Within these assumptions,  
} the effective potential in the equation for $\phi({\bf R})$ is given by
$-(N-2){\kappa^2}/{\nu}$,
where $\nu\equiv \mu_{(2\alpha)\beta}/m_\beta$ and $\kappa\equiv \kappa(R)$ 
satisfy the relation
 \begin{equation}
{\cal E}_{N-2}(R)=-(N-2)\frac{\kappa^2}{\nu},
\label{eff-pot}\end{equation}
where $\nu\equiv \mu_{(2\alpha)\beta}/m_\beta$ and $\kappa\equiv \kappa(R)$ should satisfy the relation
\begin{equation}
\left[\kappa-\frac{1}{a  }\right] R = e^{-\kappa R}.
\label{kappa}\end{equation}
The solution in the limit $a  \to\infty$ leads to
\begin{equation}
{\cal E}_{N-2}(R)=-(N-2)\frac{\gamma^2}{\nu R^2},\;\;\;{\rm where}\;\;\; \gamma=e^{-\gamma}=0.5671433.
\label{rsmall}\end{equation}
 By relaxing the unitary limit, considering any other value for $a$, the expression (\ref{kappa}) for 
 $\kappa(R)$ can be fitted within a function
\begin{equation}\label{kappa-a}
\kappa(R)\approx \frac{1}{a  }+ \left(\frac{\gamma}{R}+\frac{\varepsilon}{a  }\right)
e^{-{R}/{a  }},
\end{equation}
where the constant $\varepsilon$ is adjusted numerically. With good accuracy we obtain
$\varepsilon\equiv 0.185$. With the above expression for $\kappa(R)$, 
 the effective potential ${\cal E}_{N-2}(R)$ for the two-heavy particle system, 
\begin{equation}
{\cal E}_{N-2}(R)=-\frac{(N-2)}{\nu a^2}
\left[1+\left(\frac{\gamma a  }{R}+\varepsilon\right) e^{-\frac{R}{a  }}\right]^2,
\label{ex-e}\end{equation}
will satisfy both limits $R\ll a  $ and $R\gg a  $.
Near the unitary limit, where $R\ll a $, by keeping in the potential the next Coulomb-like term, the 
bound-state equation for $(N-2)$-light and two-heavy particles is 
{\small \begin{equation}
\left[\frac{d^2}{dR^2}  + \frac{(N-2)m_\alpha}{2\mu_{(2\alpha)\beta}}
\left(\frac{\gamma^2}{R^2}+\frac{0.7008}{R a}\right) - {\cal B}_{N} \right] u=0
,\label{srN}\end{equation}} 
where ${\cal B}_{N} \equiv -\frac{m_\alpha}{\hbar^2} E_{N} $ and
$u\equiv u(R)\equiv R \,\phi({\bf R})$. 
In the present adiabatic approximation, with $m_\alpha\gg m_\beta$, 
we have $\mu_{(2\alpha)\beta}\sim m_\beta$, such that
 $\mu_{(2\alpha)\beta}/m_\alpha$ gives approximately the light to heavy mass ratio.
In the following, the mass-ratio will be defined by  $A\equiv m_\beta/m_\alpha \ll 1$. 

The effective potential given by transcendental Eq. (\ref{kappa}) is valid close
to a shallow bound state for the light-heavy system, near the unitary limit, being more 
adequate for broad Feshbach resonance where the effective range ($R_e$) can be disregarded.
For a narrow resonance, as shown in \cite{petrov2}, one has to take into account in 
 Eq. (\ref{kappa}) the effective-range correction brought by $R^*$, such that  we should have
\begin{equation}
\left[\kappa-\frac{1}{a  }+R^*\kappa^2\right] R = e^{-\kappa R}\, ,
\label{kappa-n}\end{equation}
instead of Eq.~(\ref{kappa}). In this case, in the region $R_e\ll R\ll R^*\ll |a|$, the solution will give us
a Coulomb-like potential for the heavy-heavy system \cite{petrov2}, as
\begin{equation}
\kappa^2\sim (R\,R^*)^{-1} .
\end{equation}
However, it was also pointed out that, in region $R_e\ll R^*\ll R\ll |a|$, the adiabatic potential obtained 
from the solution of Eq. (\ref{kappa-n}) recovers the $1/R^2$ tail, which is the situation that we consider
in the following.

 For a radial potential $\Lambda/R^2$, where $\Lambda$ is dimensionless, the system has no bound-state for
 $\Lambda > -1/4$, and is anomalous for  $\Lambda < -1/4$ due to the singularity at $R\to 0$. 
 There is no lower limit in the energy spectrum, which requires a regularization, such that $R>r_1$, where $r_1$ is 
 a radial short-range cut-off. Therefore, for a boundary condition we fix the wave function to zero at $R=r_1$. 
 It is important to note that the geometric scaling property is independent on the value of $r_1$.
  In the unitary limit ($a\to\infty$),  for both broad (in the region $R_e\ll R\ll |a|$) and narrow (in the region $R_e\ll R^*\ll R\ll |a|$) 
 Feshbach resonance, we have
\begin{eqnarray}
&&\left[\frac{d^2}{dR^2} + \frac{s_{N}^2+\frac{1}{4}}{R^2}-{\cal B}_{N}
\right] u = 0 \;\; (N\ge 3),
\label{sr4}\end{eqnarray}
where $s_{N}\equiv s_{N}(A) \equiv \sqrt{\left(\frac{2+A}{4A}\right)(N-2)\gamma^2-\frac{1}{4}}$ (function of the mass ratio) is 
defining the adiabatic scaling factor. For the corresponding three-body system ( $N=3$), this scaling
factor should correspond to the non-adiabatic one, which is usually defined as $s_0$~\cite{2006braaten}.
(In the following, we take $s_3$ as defining our adiabatic value for $s_0$).

In our simplified scheme, we are generalizing the BO approach to the case of two-heavy and $(N-2)$-light bosons, 
in a way that we can obtain a general relation between the corresponding scaling factors 
with the case that we have just one-light boson:
 \begin{eqnarray}
s_{N}^2 &=& (N-2) s_{3}^2 + {(N-3)}/{4}\nonumber\\
 &\simeq& (N-2) s_{0}^2 + {(N-3)}/{4}\,\;\; (N\ge 3),
\label{sr8}\end{eqnarray}
which implies that $s_{N}>s_{N-1}$, and therefore the geometrical ratio between the energies of two successive states of the 
$N-$particle system is smaller than the corresponding ratio for the $(N-1)$-particle system. This pattern seems to persist even in the 
case where the BO approximation is not applicable like in what was found theoretically for the four and three-boson systems with 
a zero-range potential when $\mathcal{B}_4^{(1)}/\mathcal{B}_4^{(0)}\sim 1/127$, with  $\mathcal{B}_3^{(0)}<<\mathcal{B}_4^{(1)}$ 
in the strict unitary limit (for zero two-body bound-state, $B_{\alpha\beta}=0$)~\cite{2011hadizadeh}.

Therefore, the bound-state spectrum for the two-heavy and $(N-2)$-light boson, with identical particles not interacting, is 
obtained by the solution of Eq.~(\ref{sr4}), 
which follows in exact analogy with the BO approach for the three-body case, where we have two-heavy 
and one-light bosons. As detailed in Ref.~\cite{1979fonseca}, the three-body spectrum is obtained from the zeros of a modified 
Bessel function of the second kind with pure imaginary order ${\rm i}s_3$ (as defined in \cite{AS}): 
$u(R)=\sqrt{\kappa_3 R} K_{{\rm i}s_3}(\kappa_3 R)$, where $\kappa_3\equiv \sqrt{\mathcal{B}_3}$.
From the condition that the wave-function must be zero at some short distance, with a cut-off regularizing the potential at $R=r_1$,
for shallow bound-state levels, we have $\sqrt{ {\mathcal{B}^{(n)}_3}} r_1 = e^{\displaystyle{-n\pi/s_3}}\times f(s_3)$,
where $f(s_3)$ is a constant factor which does not depend on specific levels. 
From this solution, emerges the well-known geometric scaling of the three-body spectrum, with 
${\cal B}_3^{(n)}= e^{\displaystyle{-2n\pi/s_3}}{{\cal B}_3^{(0)}} \; (n=0,1,...)$, as well as the fact that the bound-state energies are
scaling with the inverse square of the cut-off at short distances, $1/r_1^2$. 

We should also note that, the boundary condition of the wave-function at long distances is giving by the absolute value of the 
two-body scattering length, with the number of the levels in the spectrum being 
\begin{equation}
{\cal N}_3 \simeq \frac{s_3}{\pi}{\rm ln}(|a|/r_1),
\label{N3} \end{equation} which is infinite in the unitary limit~\cite{1979fonseca}.
As we move away from the unitary limit, the number of trimers decrease with the ratio between adjacent binding 
energies following a scaling relation, as shown in Fig. 2 of Ref.~\cite{1999-pra60-R9} for the case of three-identical particles.

Before going to the next section where our aim is to analyze the inter-relation between the spectrum of a $N-$boson system
with the spectrum of subsystems, it is of interest to check the extension of the validity of the adiabatic Born-Oppenheimer approach, 
close to unitary limit. For that, we are verifying numerically the $s_3$ values, obtained for the case with $N=3$ (one light and 
two boson system) for different values of the mass-ratio $A\equiv m_\beta/m_\alpha\ll 1$,
in comparison with the values of $s_0$ reported in Ref.~\cite{2006braaten}.   
The results presented in Table~\ref{tab1} are illustrative on the accuracy of the BO approach, which improves as the mass ratio $A$
decreases. 

\begin{table}[tbh!]
\caption{Values of the scaling factor $s_3$ and $e^{{\pi}/{s_3}}$, obtained by solving the adiabatic equation
(\ref{sr4}) in comparison with the respective exact values as reported in Ref.~\cite{2006braaten}. 
}
 \begin{tabular}{c|cccccccc}
 \hline\hline
$A$          &0.1        &0.05      &0.04        &0.03        &0.02        &0.01        &0.001\\
\hline\hline
$s_3$           &1.1995&1.7456&1.9624&2.2784&2.8057&3.9891&12.675\\
$s_0$           &1.4682&1.9194&2.1142  &2.4067&2.9084  &4.0612  &12.698\\
$e^{{\pi}/{s_3}}$&13.725&6.0483&4.9574&3.9703&3.0641&2.1980&1.2813\\
$e^{{\pi}/{s_0}}$&8.4977&5.1383&4.4193 &3.6889&2.9452 &2.1675&1.2807\\
\hline\hline
  \end{tabular}
\label{tab1}\end{table}

\subsection{Two-heavy and two-light bosons}
As discussed in the previous subsection,   
the solutions for the spectrum of two-heavy and $(N-2)$-light bosons are obtained by
following in close analogy  the same analytical expression as in the case of $N=3$.
Therefore, the bound-state wave functions presented in Eq.~(\ref{sr4}) are 
given by modified Bessel functions of the third kind with pure imaginary order  
${\rm i}s_{N}$, such that $u(R)=\sqrt{\kappa_N R} K_{{\rm i}s_N}(\kappa_N R)$, with
$\kappa_N^2\equiv {\cal B}_N$.
However, the cases with $N\ge 4$ will differ from the case of $N=3$ by the boundary conditions.
For example, in the case that $N=4$ (two-heavy and two-light bosons), with the wave function vanishing 
at $R=r_2$, the shallow energy states in the spectrum are given by 
$\sqrt{{\mathcal{B}^{(n)}_4}} = \left(r_2^{-1}e^{\displaystyle{-n\pi/s_4}}\right)\times f(s_4)$.
Let us emphasize that the condition of having the wave-function vanishing at $R=r_2$
represents the information associated to a short-range four-body scale.

At the short range, the cut-off of the long-range Efimov-like potential is associated to a four-body 
 short-range parameter; and, at the long range, associated to the size of the three-body system.
In analogy with the three-body case, in which the spectrum is restricted by the size of the two-body bound-state,
the four-body spectrum is restricted by the size of the three-body level that we are considering. 
 Following this physically motivated picture, in the case of four-body system, one of the heavy particles 
 can only probe the long-range potential if it is in a region within the tail of the remaining three-body bound state. 
 If it is more distant it cannot interact with the other heavy particle through the Efimov-like long-range potential. 
Therefore, similar as in the three-body case, where the number of leves is given by Eq.~(\ref{N3}),
for the expected number of four-body levels attached to the three-body ground-state level we obtain  
\begin{equation}
{\cal N}_4^{(0)} \simeq -{\displaystyle \frac{s_4}{\pi}}{\rm ln}\left(\sqrt{B^{(0)}_3} r_2\right),
\label{N4levels}
\end{equation}
which shows that, by increasing a given three-body binding energy, the corresponding number can change according 
to this relation. 
Similar as in the three-body case, the ratio between adjacent tetramer energies will follow a scaling 
relation. This scaling relation was verified in Ref.~\cite{2011hadizadeh}, by solving the full FY four-body 
system, considering identical particles.

Therefore, the realization of a maximum (infinite) number of tetramers is possible only if the trimer spectrum is collapsed in the 
ground state with zero bound-state energy. Otherwise, the maximum number is restricted by the size of the ground-state trimer.

In order to proceed, we see here the convenience to replace the previously mentioned labels $n$ of the three-body 
Efimov spectrum, by $n_3$, with two-labels defining the possible four-body spectra ($n_4$ and $n_3$), due to the 
fact that for each three-body level we can possible have a four-body spectrum.
Therefore, the geometric scaling of the three-body spectrum is given by
{\small\begin{equation}{\cal B}_3^{(n_3)}= 
e^{\displaystyle{-2n_3\pi/s_0}}{{\cal B}_3^{(0)}},\;(n_3=0,1,...),\;\; {{\cal B}_3^{(0)}}\sim r_1^{-2},
\label{eq11}\end{equation}} 
The factor $s_3$ was replaced by the exactly known values $s_0$, as given in Table~\ref{tab1}, in order to improve 
the approximation of our results obtained for the cases we have four or more particles.
Therefore, from Eq.~(\ref{sr8}), 
the relation between trimer ($N=3$) and tetramer ($N=4$) scaling factors, is given by
$s_{4}^2 = 2 s_{0}^2 + {1}/{4}$.
So, for a given level $n_3$, we have the following relation for the four-body  levels:   
\begin{equation}
{\cal B}_{4,n_3}^{(n_4)}= e^{\displaystyle{-2\,n_4\pi/s_4}}{{\cal B}_{4,n_3}^{(0)}}.
\label{eq18}\end{equation}
This is realizable for the four-body states below the ground-state trimer ($n_3=0$), being limited in the other
cases, as will be discussed.
 
The BO relation given by Eq.~(\ref{sr8}), for the specific case of $N=4$ with two heavy and two light bosons, was also 
presented recently in Ref.~\cite{2018naidonFBS}, under the same simplified conditions where only the non-identical particles 
(heavy-light) have non-zero interactions. By considering $A\ll 1$ the BO approximation was shown to be fully consistent 
with non-adiabatic FY calculations.

\section{Interwoven cycles}

\subsection{Three and four-body spectrum}\label{subsec3+4body}
\begin{figure}
  \includegraphics[width=\linewidth]{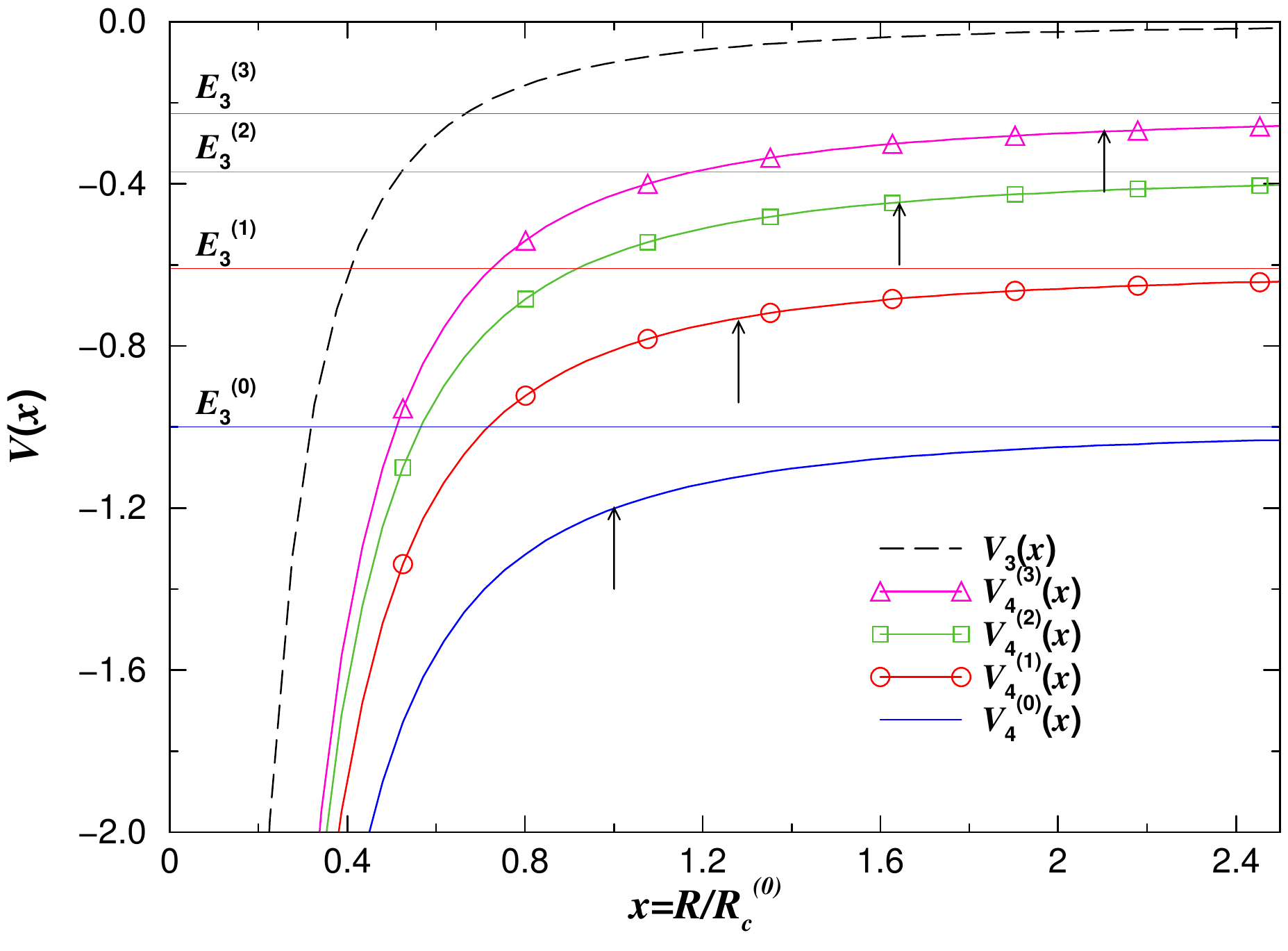}
  \caption{The effective long-range potential (in the unitary limit) 
  between the two heavy particles, in the BO approximation of the two-light and two-heavy particle system, 
considering the different 3+1 thresholds. The potential and all the energies are dimensionless, scaled by a 
factor 100 times the three-body ground-state energy, with $x$ given in terms of the long-range cut-off 
$R_c^{(0)}$, where the four-body system is damped at the size of the three-body ground-state. 
  The horizontal dashed lines indicates the 3+1 dissociation  threshold at the three-body
energies $E_{3}^{(n)}$ ($n=$0 is the ground state), with the ratio between them fixed by the Efimov factor $e^{2\pi/s_0}$.   
The arrows indicate schematically  where the long-range potential in the 4-body system is damped, 
at the size of the $n-$th 3-body state. 
}.
  \label{fig1}
\end{figure}

Notice that, by considering a tetramer, with two-heavy and two-light particles, where only the
light-heavy particles interact weakly (such that $B_{\alpha\beta}$ is close to zero), we should 
have a four-body spectrum  ($\alpha\alpha\beta\beta$) interconnected with two identical three-body 
spectrum ($\alpha\alpha\beta$). We have the energy ratios for the trimer and the tetramer spectrum from the tail
of the long-range potential. But we should point out the relation between the four-body and three-body 
levels. Concerning that, we have the schematic Fig.~\ref{fig1} to illustrate the dependences of the tetramers 
on the trimer energies.

The effective BO potential for the two heavy particles, at large distances has the three-plus-one (3+1)  
channel threshold, for each possible trimer state, as shown in the figure. The effective potential holds the 
tetramer bound states below the ground state trimer, otherwise tetramer resonances are placed in the 
effective potential below each excited trimer. 
The size of each trimer cuts down the long-range effective BO potential in the tetramer system (indicated
in the figure by the arrows), and therefore when it is decreased, or the trimer binding increases, the excited 
tetramers tends to disappear in the 3+1 threshold as the attraction drops. However, the 
correlation between the energies of successive tetramers in the potential pocket is not destroyed due to
the universal long-range potential (well known in the Efimov physics). 
This correlation was indeed verified in the case of four identical bosons at the unitary limit \cite{2011hadizadeh}. 
Besides that, no matter to which trimer they are associated, the correlation between successive tetramer levels 
is verified to be universal and again dominated by a long-range potential, with strength larger than the 
corresponding one for the trimer.

In the case that $A=0.01$, from the adiabatic results given in Table~\ref{tab1}, we obtain 
${{\cal B}_3^{(n_3)}}/{{\cal B}_3^{(n_3+1)}}=$4.8312, implying that 
${{\cal B}_{4,n_3}^{(n_4)}}/{{\cal B}_{4,n_3}^{(n_4+1)}}=3.0326$.
However, as explained before we can improve the results obtained for the case of $N\ge 4$, by considering the 
non-adiabatic results obtained for the three-body scaling factor. In this way, we have
 ${{\cal B}_3^{(n_3)}}/{{\cal B}_3^{(n_3+1)}}=4.6979$
and ${{\cal B}_{4,n_3}^{(n_4)}}/{{\cal B}_{4,n_3}^{(n_4+1)}}=2.9739$. 
The Table~\ref{tab2} displays our results up to $N=6$, for mass-ratios $A$ between 0.001 and 0.04.

\begin{table}[tbh!]
\caption{In this table, we are applying the expression (\ref{sr8}) to obtain the corresponding 
energy ratios in the spectrum of two-heavy and $(N-2)$-light particles, when the interaction is restricted
to inter-species atoms. For that, we are  considering the 
non-adiabatic scaling factor for the three-body system. }
\begin{tabular}{c|cccccc}
 \hline\hline
$A$                               &0.04        &0.03        &0.02        &0.01        &0.001\\
\hline\hline
$s_{0}$                              
&2.114&2.407&2.908&4.061&12.70\\
$e^{\displaystyle{2\,\pi/s_0}}$
&19.5  &13.6  &8.67  &4.70  &1.64\\   
$e^{\displaystyle{2\,\pi/s_4}}$   
&7.95  &6.21  &4.57  &2.97  &1.42\\
$e^{\displaystyle{2\,\pi/s_5}}$    
&5.39  &4.42  &3.44  &2.43  &1.33\\
$e^{\displaystyle{2\,\pi/s_6}}$   
&4.29  &3.61  &2.91  &2.16  &1.28\\
\hline\hline
  \end{tabular}
\label{tab2} 
\end{table}

Now, we should note that the above relations are giving us the ratio between two consecutive states of the 
spectrum for a fixed number of $N-$ bosons. However, each spectrum of a given number $N-$light boson
should be related to the spectrum of $(N-1)$ bosons. 
All these spectra are related to the light-heavy unitary limit, which is $B_{\alpha\beta}=0$ (or $a_{\alpha\beta}\to\infty$), 
such that, we have the spectra given by
interwoven limit cycles. 
Let us consider explicitly the bound-state spectra (negative energies), with a maximum 
of four particles, in the unitary limit (${{\cal B}_2}=0$), where  the three-body levels are given by 
$E_3^{(n_3)}\equiv -{{\cal B}_3^{(n_3)}}$ and, for each three-body level $n_3$ we have the corresponding
four-body levels $E_{4,n_3}^{(n_4)}\equiv -{{\cal B}_{4,n_3}^{(n_4)}}$. In this case, the spectra should be
as follows:
\begin{small}
 \begin{eqnarray} 
B_3^{(0)}&>&e^{-2\pi/s_0}B_3^{(0)}>e^{-4\pi/s_0}B_3^{(0)}>...>0;\nonumber\\
B_{4,0}^{(0)}&>&e^{-2\pi/s_4}B_{4,0}^{(0)}>e^{-4\pi/s_4}B_{4,0}^{(0)}>...> B_3^{(0)};\nonumber\\
B_{4,j}^{(0)}&>&e^{-2\pi/s_4}B_{4,j}^{(0)}>e^{-4\pi/s_4}B_{4,j}^{(0)}>...>e^{-2j\pi/s_0}B_3^{(0)};\nonumber
\end{eqnarray}
\end{small}
So, it is clear that the number of states between two three-body states are limited by the corresponding
ratio between three-body levels.

From the relation between the scaling factors $s_4$ and $s_0$, given by Eq.~(\ref{sr8}),
together with our adiabatic results obtained in Table~\ref{tab1} for the trimers, one can easily conclude that 
no more than one tetramer related to $B_3^{(1)}$ can exist with bound-state energy 
between $B_3^{(0)}$ and $B_3^{(1)}$, for $A<<1$. 
This picture will change according to the mass ratio $A$, with the possibility of at least one more tetramer level 
appearing between two trimers for $A\approx 1$, which is consistent with verified experimental results.

To illustrate  this last point, let us consider, for example,  the tetramers between the 
ground and 1st excited trimer levels for $A=0.01$. Now, suppose that one tetramer resonance within these two 
trimer levels has an  energy close to the ground state trimer, then according to Table~\ref{tab2}, the next tetramer resonance
following the geometrical ratio will have an energy of $B_3^{(0)}/2.97$, while $B_3^{(1)}=B_3^{(0)}/4.70$. 
This suggests that only one tetramer is possible for strong mass imbalanced systems. For identical bosons, at the unitary limit, 
 up to three tetramers levels are possible to exist 
between two Efimov trimers\cite{2011hadizadeh}. Indeed, Table~\ref{tab2} shows that by decreasing $A$, the difference 
between the trimer and tetramer geometrical ratios increases allowing eventually more tetramers between two successive trimers.

The existence of an  Efimov potential for two heavy bosons (see Fig.~\ref{fig1}) in a
four-body system different from the one  in the heavy-heavy-light system
justifies the independent universal correlation found for the energies of two successive tetramers in 
the four-boson system with a zero-range interaction. In this case  a
 four-body scale is necessary even when the two- and three-boson scales are fixed. The new limit cycle that
 was theoretically obtained in ~\cite{2011hadizadeh} for the four-boson system, has its counterpart in the
 heavy-heavy-light-light system: it is a manifestation of the new  effective 
  Efimov potential for the four-body system independent of the three-body one. When for example, a short
  range four-body force is varied, the heavy-heavy-light-light bound states  will follow an universal correlation curve, namely a new limit cycle, which generalizes what was found for the 
 four-boson system in ~\cite{2011hadizadeh}. Such a situation can be potentially found in cold atomic gases
when narrow Feshbach resonances are present in the atom-atom system, and effective short-range 
few-body forces in the open channel are active, and may allow to disentangle the trimer and tetramer binding energies, 
beyond the van der Waals universality.
See Ref.~\cite{2018bazak}, for a recent effective-field theory approach related to the four-body scale, supporting
previous predictions~\cite{YamaEPL2006,2011hadizadeh} of an independent four-body scale.

A new limit cycle that  is found in a four-body $\alpha\alpha\beta\beta$ system emerges from the different 
Efimov long-range potential, which provides a new value for the  geometrical ratio between the tetramer energies in the limit where the
size of the $\alpha\alpha\beta$ trimer goes to infinity. As discussed before, the tail of the potential is cut at the size of the subsystem.
To provide a concrete example of an independent limit cycle, we study the universal scaling function defined  in the unitary limit by
\begin{equation}\label{scal}
 {\cal B}^{(n+1)}_{4,i}/{\cal B}^{(n+2)}_{4,i}= {\mathcal G}_4 \left({\cal B}^{(n)}_{4,i}/{\cal B}^{(n+1)}_{4,i}\right)\, ,
\end{equation} 
where $n$ are defining the four-body levels, with $i$ defining the three-body levels. 
This relation (\ref{scal}) expresses the universal correlation between the binding energy ratios of successive $\alpha\alpha\beta\beta$ states,
when the $\alpha\alpha\beta$ energy is changed. Note that the dependence on $B_3^{(i)}$ is implicitly accounted in the ratio 
${\cal B}^{(n)}_{4,i}/{\cal B}^{(n+1)}_{4,i}$, as the general scaling proposed in \cite{2011hadizadeh} correlates,
${\cal B}^{(n+1)}_{4,i}/{\cal B}^{(n)}_{4,i}$ with ${\cal B}^{(i)}_{3}/{\cal B}^{(n)}_{4,i}$.

\begin{figure*}
\includegraphics[width=5.8cm]{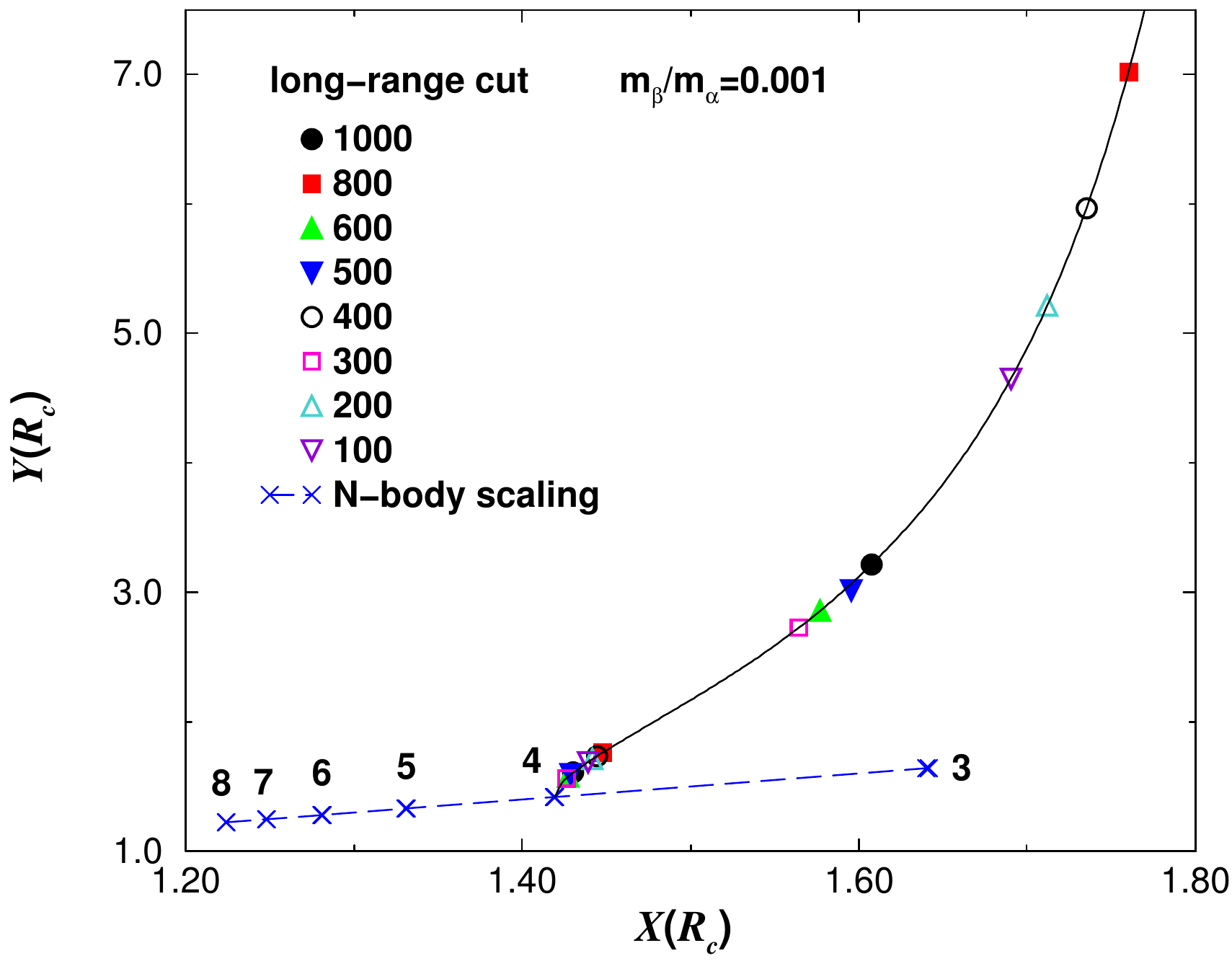}
  \includegraphics[width=5.5cm]{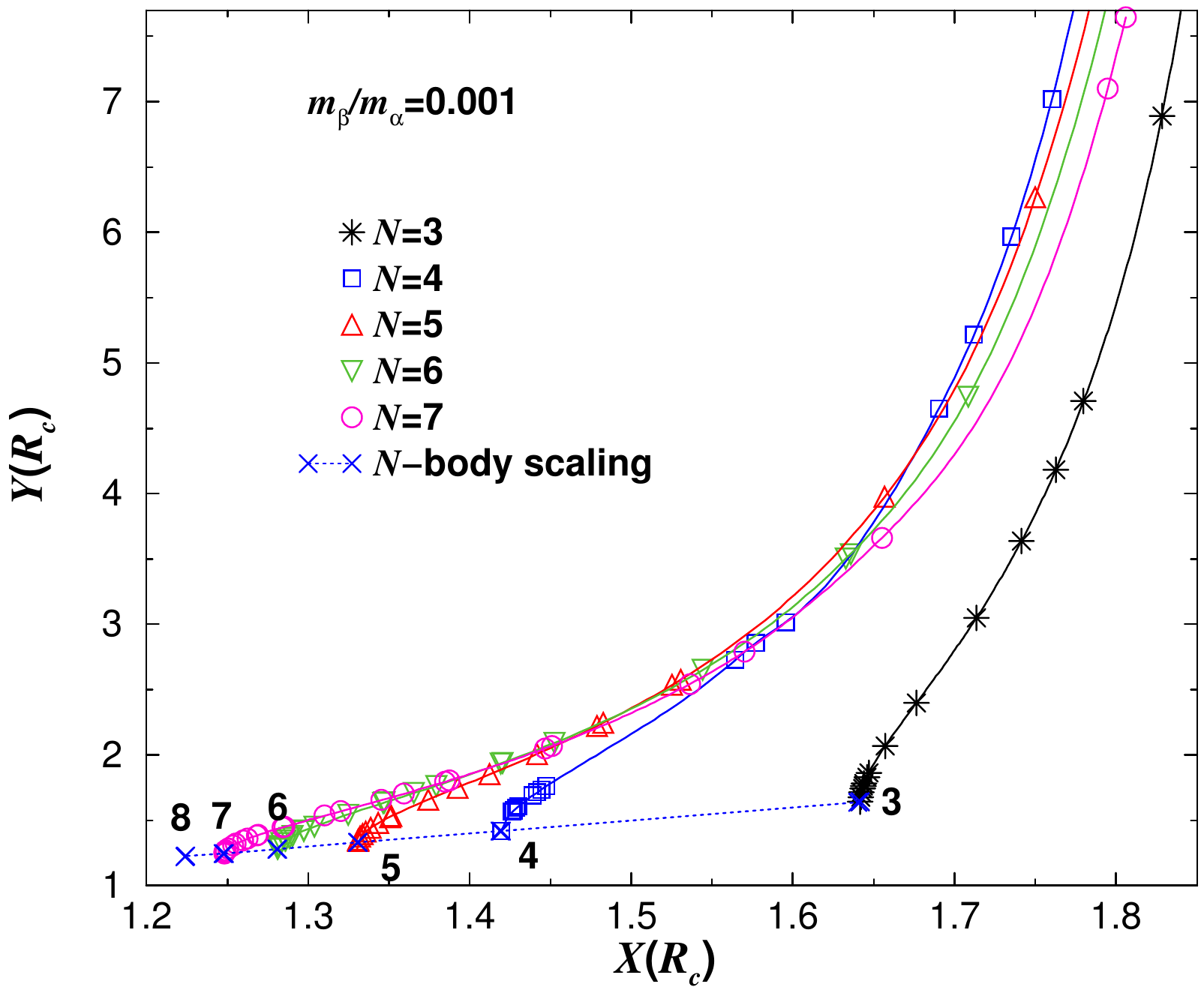}
  \includegraphics[width=5.5cm]{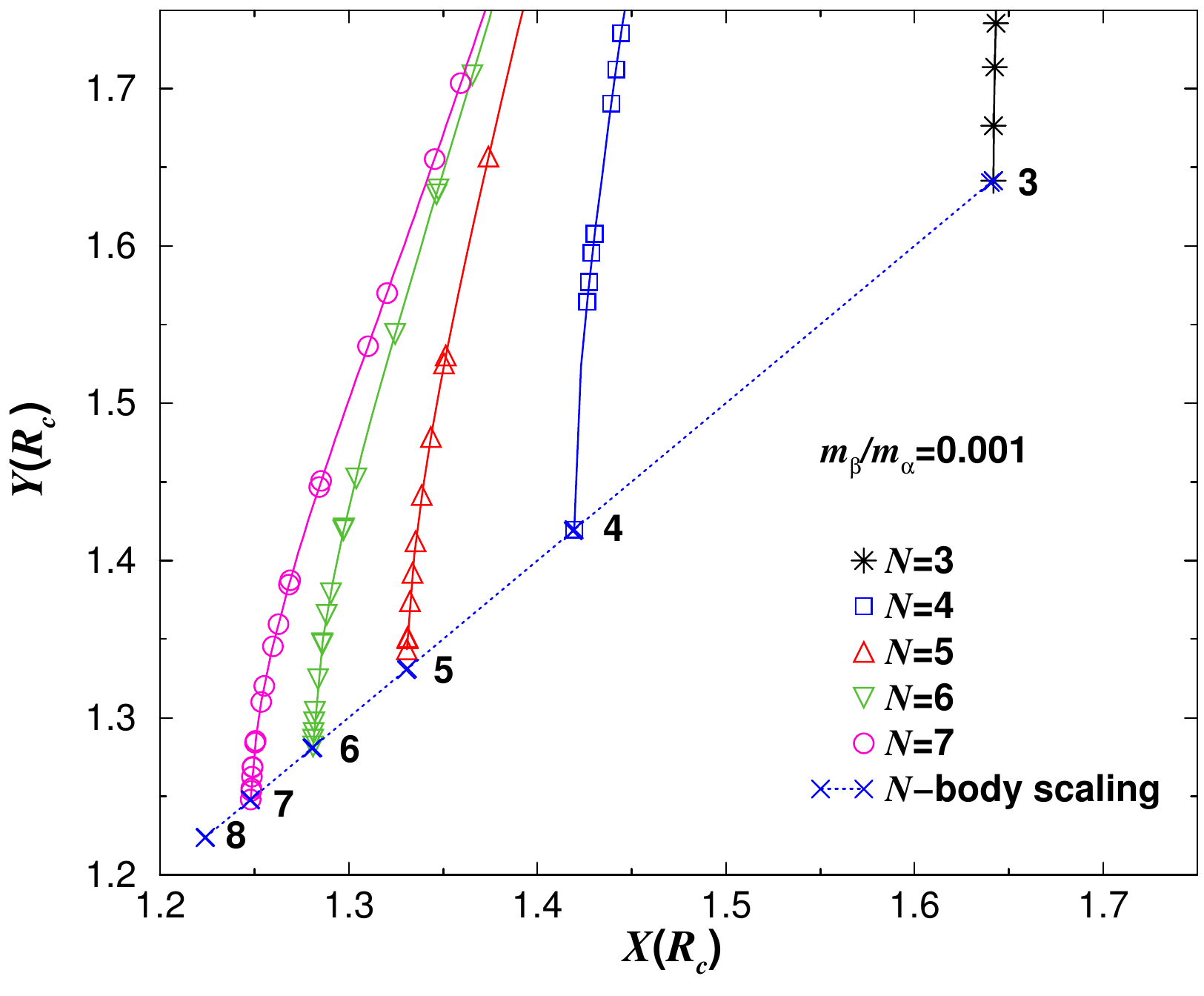}
    \caption{
Scaling plots corresponding to Eqs.~(\ref{scal}) and (\ref{scalK}), for $A=1/1000$
with $s_N$ given Eq.~(\ref{sr8}) and considering 
the definitions $X(R_c)\equiv  {\cal B}^{(n)}_{N,i}/{\cal B}^{(n+1)}_{N,i}$ and  
$Y(R_c)\equiv {\mathcal G}_N \left(X\right) =   {\cal B}^{(n+1)}_{N,i}/{\cal B}^{(n+2)}_{N,i}$,
where $R_c\equiv R^{(N-1)}_{c,i}$ is the long-range cut of the level $i$ of$(N-1)-$body wave-function.
In the left panel, we consider $N=4$, in which the dashed-blue line we have the analytical limit given by $Y=X$, 
  where we are indicating the total number of bosons (from 3 to 8). The solid-black line shows our results for the case 
  with $N=4$, with the symbols corresponding to specific long-range cuts, as indicated inside the figure.
In the middle and right panels, $N$ is an arbitrary number between 3 and 7. 
  With the dotted-blue line we are connecting the analytical limits   ($Y=X$) for $N=$3 to 8 (as indicated).
By varying $R_c$, for each $N-$boson system we verify that the corresponding curves (represented by solid lines with symbols
indicated inside the panels) are converging to the corresponding analytical limit.  
In the lower panel we present a view to the region very close to the analytical limit. }
\label{fig2}
\end{figure*}

The universal function $ {\mathcal G}_4$ obtained in the unitary limit does not depend on $n$. When a given trimer level $n$ has
its size close to infinity, the state explores the long-range $1/R^2$ behavior of the potential, with the corresponding function 
converging to the analytical scaling limit, such that
${\mathcal G}_4 \left(e^{2\pi/s_4}\right)=e^{2\pi/s_4}$. 
 In Fig.~\ref{fig2}, for this case that we have two-light and two-heavy boson with a mass-ratio $A=m_\beta/m_\alpha=$1/1000, 
 we show the scaling behavior corresponding to the function (\ref{scal}) as we vary the long-range cut-off $R=R_c$ of the 
 wave-function.  For that, we solve Eq. (\ref{sr4}) for $N=4$, by considering the scaling factor 
 $s_4=17.965$, which is obtained from Eq.~(\ref{sr8}) once $s_3$ is replaced by $s_0=12.698$ (as given in Table~\ref{tab1}).
 The treatment to solve the Eq. (\ref{sr4}) was followed by considering an analytical reduction to a transcendental equation,
 which is solved numerically. For the boundary conditions, we consider an infinite barrier fixed at $R=r_2=1$, with the long-range 
 cut-off varying from $R_c=$100 to 1000. The results are indicated inside the figure, where for each cut-off we obtain 
 three or four energy levels.
 This scaling function can only be fully realized for tetramers below the ground state trimer, as we have showed that for such large 
 mass asymmetries no more than two tetramers are possible between two Efimov trimers.
 As shown, by varying the long-range cut-off, we observe the convergence of the results to the expected scaling limit.

\subsection{More than four bodies }\label{subsecmorebodies}

Let us consider a generalization of the previous discussion addressing the three and four-body systems to more 
than four particles.
We start with the five-body problem, with two-heavy and three-light bosons, where the interactions are only between
heavy and light particles close to the unitary limit (as before). In this case, we have the $4+1$ and the $3+2$ thresholds: 
The five-body states that are bound will have the long range potential damped at the size of the four body state; as well as 
affected by the existing three-body bound state levels.
 Therefore, among the sub-system states, the question is to find out which one has the smallest size and cuts the 
 long-range effective five-body  potential. 
 Nevertheless, the physical picture has to be extended for the five-body states when three and four-body excited states exist. 
 Here, we still remain with the problem of finding the relevant threshold that determines the asymptotic value of the potential. 
 To be simple, assuming that the $4+1$ threshold is the relevant one and therefore the energy of the excited four-body state 
 defines the threshold for the five-body resonances that are formed in the $4+1$ scattering channel. We may ask what about the
 $3+2$ channel? Certainly the size of the two bound states, namely the two and three-body bound states, 
 are larger than the four-body bound state, and they  are less effective to cut the tail of the long range potential. 

The discussion goes further as we add more light bosons to the system.  Then, more scattering channels are possible; becoming
apparently more difficult to identify which are the ones more effective in providing the range. 
However, one situation that can be considered more likely of being generalized is when we add one light boson 
to a $N-$body system which is in a Borromean state, considering that only one two-body scattering threshold exists, namely, the 
$(N-1)+1$ threshold. 
In this case, the interwoven spectrum is analogous to the one described for the trimer-tetramer case. 
 
We illustrate the evolution of the states when the long-range cut-off is moved by considering the scaling function for the $N-$body
system [two-heavy and $(N-2)$-light bosons], analogous to the one written for the tetramer case (\ref{scal}).
In the unitary limit ($B_{\alpha\beta}=0$), the scaling between the $N-$body energy levels, as one varies the corresponding 
long-range cut-off {  $R_{c,i}^{(N-1)}$ (where $i$ refer to the levels of the $(N-1)-$system)
is defined by
\begin{equation}\label{scalK}
 \frac{{\cal B}^{(n+1)}_{N,i}(R_{c,i}^{(N-1)})}{{\cal B}^{(n+2)}_{N,i}(R_{c,i}^{(N-1)})}= 
 {\mathcal G}_N \left(\frac{{\cal B}^{(n)}_{N,i}(R_{c,i}^{(N-1)})}
 {{\cal B}^{(n+1)}_{N,i}(R_{c,i}^{(N-1)})}
 \right)\;\;(N\ge 4),
\end{equation} 
with $n$ labeling the energy level of the $N-$body system.}
This relation expresses the universal correlation between the binding energy ratios of successive states for two-heavy-boson systems
with $(N-2)$ and with $(N-3)$ light bosons.
Note that the dependences on the other scales are wiped out as we assume that in this example the $N-2$, $N-3$, ...  are much 
larger than the $N-1$ system. This can be regarded as a situation close to a Brunnian system \cite{YamaPRA2010}. 
Corresponding to the above described conditions, in the middle panel of Fig.~\ref{fig2}, we present our full results for the 
correlations obtained for $N=$ 3 to 7, considering the mass ratio $A=0.001$, where we demonstrate that the scaling is converging to the 
analytical expression given by ${\cal G}_N\left(e^{2\pi/s_N}\right)=e^{2\pi/s_N}$. A closer look to the limit is given in the right panel of
this figure. The patterns are the same as in the tetramer case, and the only difference is the Efimov factor which determines the strength 
of the long-range potential.

\section{Summary}
Firstly, let us summarize the relevant physical aspects of the three and four-body 
interwoven energy spectra presented in the discussion of Sect. \ref{subsec3+4body}:
\begin{enumerate}[(i)]
\item The three-body thresholds for the ground and excited states 
give the asymptotic value of the four-body long range effective interaction. 
\item The separation between the asymptotic values of the long range effective 
potential follows the geometrical three-body energy ratio in the unitary limit, and 
provides the thresholds for the four-body states attached to the light-heavy-heavy trimers.
\item The effective four-body long-range potential itself carries a proper geometrical 
scale different from the three-body one, $s_4 > s_3 $ (where $s_3=s_0$), exhibiting  a proper limit cycle independent 
of the three-body one, which is damped at the size of the trimer in the unitary limit.
\item The scaling function correlating the ratios between two close tetramer levels follows the same trend as in the three-body Efimov
case, but with a different scaling factor $s_4$.
\item For very-large mass asymmetry as considered in the present BO analysis, the ratio between the energies of the trimer levels is not 
much larger than the ratio between tetramer levels, such that we have no room for more than one tetramer level between two trimer levels. 
\end{enumerate}
Secondly,  the discussion on Sect.\ref{subsecmorebodies} for more than four body systems, namely 
two-heavy and $(N-2)-$light particles, becomes more complex due to the different scattering thresholds. 
However, some general properties can also be summarized as:
\begin{enumerate}[(i)]
\item The effective interaction between the two-heavy bosons has an Efimov-type potential at unitarity, with strength increasing 
linearly with the number of  light particles and  a correspondingly decreasing  geometrical separation between the bound states.
\item The effective $N$-body long-range potential asymptotically goes to the lowest scattering threshold (see Fig.~\ref{fig1}). For 
the simplest situation where only the $(N-1)-$state is bound, it corresponds to its binding energy.
\item The heavy-heavy effective interaction for the $N$-body system is damped at the
size of bound-state levels of the $(N-1)$-system. This determines the maximal number of possible weakly bound states composed by  $(N-2)$-light
boson and two heavy ones.
\end{enumerate}

In order to verify the predicted interwoven spectrum in an experimental realization we suggest to consider ultracold quantum 
mixtures of two atomic species with strong mass asymmetry, such as the systems which are being investigated with  
yterbium and lithium~\cite{2011hansen}, as well as mixtures with other alkaline-earth atoms~\cite{2011hara,2015makrides}.
In experiments with a given set of atoms, different narrow Feshbach resonances have to be exploited for the inter-species 
 in order to control the two-body scattering length, such that the weakly-bound  $N-$body systems close to the $N-1$ 
threshold ($N\ge 4$) near to the unitarity ($|a|\gg R^*$) will have sizes much larger than $R^*$. 
In this way, as an example, it should be possible to explore the correlations between the positions of the recombination 
resonances  coming from two successive  tetramer bound states crossing the continuum threshold,
which will move along the correlation curve and break the van der Waals universality. 
All these exciting possibilities also demand further theoretical and experimental efforts in order to determine the characteristics 
of these induced effective few-atom forces in the open channels close to  narrow Feshbach resonances.

\acknowledgements
The authors thank the following agencies for partial support: 
Conselho Nacional de Desenvolvimento Cient\'\i fico e Tecnol\'ogico
[Procs.310205/2017-4(AD), 308486/2015-3(TF), 306191-2014-8(LT) and
INCT-FNA Proc.464898/2014-5], 
Funda\c c\~ao de Amparo \`a Pesquisa do Estado de S\~ao Paulo 
[Projs. 2017/05660-0(LT and TF)], 
and Coordena\c c\~ao de Aperfei\c coamento de Pessoal de N\'\i vel Superior (LT).

\end{document}